\documentclass[preprintnumbers,prd,twocolumn,showpacs,floatfix,preprintnumbers,letterpaper,amsmath,amssymb,nofootinbib,superscriptaddress]{revtex4}
\usepackage{graphicx}
\usepackage{epsfig}
\usepackage{bm}
\usepackage{amsfonts}

\newcommand{\la}{\lambda}

\newcommand{\bear}{\begin{eqnarray}}

\newcommand{\eear}{\end{eqnarray}}

\newbox\pippobox

\def\6{\partial}

\def\a{\alpha}

\def\sq
\def\a{\alpha}

\def\dn{\Delta N_{\nu}}

\newcommand {\lla} {\ {\raise-.5ex\hbox{$\buildrel<\over\sim$}}\ }

\def\be{\begin{equation}}
\def\ee{\end{equation}}
\def\ba{\begin{eqnarray}}
\def\ea{\end{eqnarray}}
\def\w{\omega}
\renewcommand{\(}{\left(}
\renewcommand{\)}{\right)}
\renewcommand{\[}{\left[}
\renewcommand{\]}{\right]}

\begin{document}

\title{Observational constraints on Ho\v{r}ava-Lifshitz cosmology}

\author{Sourish Dutta}
\email{sourish.d@gmail.com} \affiliation{Department of Physics and
Astronomy, Vanderbilt University, Nashville, TN  ~~37235}

\author{Emmanuel N. Saridakis }
\email{msaridak@phys.uoa.gr} \affiliation{Department of Physics,
University of Athens, GR-15771 Athens, Greece}

\begin{abstract}
We use observational data from  Type Ia Supernovae (SNIa), Baryon
Acoustic Oscillations (BAO), and Cosmic Microwave Background
(CMB), along with requirements of Big Bang Nucleosynthesis (BBN),
to constrain the cosmological scenarios governed by
Ho\v{r}ava-Lifshitz gravity. We consider both the detailed and
non-detailed balance versions of the gravitational sector, and we
include the matter and radiation sectors. We conclude that the
detailed-balance scenario cannot be ruled out from the
observational point of view, however the corresponding likelihood
contours impose tight constraints on the involved parameters. The
scenario beyond detailed balance is compatible with observational
data, and we present the corresponding stringent constraints and
contour-plots of the parameters. Although this analysis indicates
that Ho\v{r}ava-Lifshitz cosmology can be compatible with
observations, it does not enlighten the discussion about its
possible conceptual and theoretical problems.
\end{abstract}

\date{\today}


\maketitle

\section{Introduction}

Recently, a power-counting renormalizable, ultra-violet (UV)
complete theory of gravity was proposed by Ho\v{r}ava in
\cite{hor2,hor1,hor3,hor4}. Although presenting an infrared (IR)
fixed point, namely General Relativity, in the  UV the theory
possesses a fixed point with an anisotropic, Lifshitz scaling
between time and space. Due to these novel features, there has
been a large amount of effort in examining and extending the
properties of the theory itself
\cite{Volovik:2009av,Cai:2009ar,Cai:2009dx,Orlando:2009en,Nishioka:2009iq,Konoplya:2009ig,Charmousis:2009tc,Li:2009bg,Visser:2009fg,Sotiriou:2009gy,
Sotiriou:2009bx,Germani:2009yt,Chen:2009bu,Chen:2009ka,Shu:2009gc,Bogdanos:2009uj,Kluson:2009rk,Afshordi:2009tt,Myung:2009ur,Alexandre:2009sy,
Blas:2009qj,Capasso:2009fh,Chen:2009vu,Kluson:2009xx}.
Additionally, application of Ho\v{r}ava-Lifshitz gravity as a
cosmological framework gives rise to Ho\v{r}ava-Lifshitz
cosmology, which proves to lead to interesting behavior
\cite{Calcagni:2009ar,Kiritsis:2009sh}. In particular, one can
examine specific solution subclasses
\cite{Lu:2009em,Nastase:2009nk,Colgain:2009fe,Ghodsi:2009rv,Minamitsuji:2009ii,Ghodsi:2009zi,Wu:2009ah,Cho:2009fc,Boehmer:2009yz,Momeni:2009au},
the phase-space behavior \cite{Carloni:2009jc,Leon:2009rc}, the
gravitational wave production
\cite{Mukohyama:2009zs,Takahashi:2009wc,Koh:2009cy,Park:2009gf,Park:2009hg,Myung:2009ug},
the perturbation spectrum
\cite{Mukohyama:2009gg,Piao:2009ax,Gao:2009bx,Chen:2009jr,Gao:2009ht,Cai:2009hc,Wang:2009yz,Kobayashi:2009hh,Wang:2009azb},
the matter bounce
\cite{Brandenberger:2009yt,Brandenberger:2009ic,Cai:2009in,Suyama:2009vy},
the black hole properties
\cite{Danielsson:2009gi,Cai:2009pe,Myung:2009dc,Kehagias:2009is,Mann:2009yx,Bertoldi:2009vn,Park:2009zra,Castillo:2009ci,BottaCantcheff:2009mp,Lee:2009rm,Varghese:2009xm,Kiritsis:2009rx},
the dark energy phenomenology
\cite{Saridakis:2009bv,Park:2009zr,Appignani:2009dy,Setare:2009vm},
the astrophysical phenomenology
\cite{Kim:2009dq,Harko:2009qr,Iorio:2009qx,Iorio:2009ek}, the
thermodynamic properties \cite{Wang:2009rw,Cai:2009qs,Cai:2009ph}
etc. However, despite this extended research, there are still many
ambiguities if Ho\v{r}ava-Lifshitz gravity is reliable and capable
of a successful description of the gravitational background of our
world, as well as of the cosmological behavior of the universe
\cite{Charmousis:2009tc,Li:2009bg,Sotiriou:2009bx,Bogdanos:2009uj,Koyama:2009hc,Papazoglou:2009fj}.

Although the discussion about the foundations and the possible
conceptual and phenomenological problems of Ho\v{r}ava-Lifshitz
gravity and cosmology is still open in the literature, it is worth
investigating in a systematic way the constraints imposed by
observations in a universe governed by Ho\v{r}ava gravity. Thus,
in the present work we use Big Bang Nucleosynthesis conditions,
together  with Type Ia Supernovae (SNIa), Baryon Acoustic
Oscillations (BAO) and Cosmic Microwave Background (CMB) data, in
order to construct the corresponding probability contour-plots for
the parameters of the theory. Furthermore, in order to be general
and model-independent, we perform our analysis with and without
the detailed-balance condition. As we will show, both the
detailed-balance and beyond-detailed-balance formulations are
compatible with observations, however under tight constraints  on
the model parameters.

The paper is organized as follows: In section \ref{model} we
present the basic ingredients of Ho\v{r}ava-Lifshitz cosmology,
extracting the Friedmann equations, and describing the dark matter
and dark energy dynamics. In section \ref{Observational
constraints} we constrain both the detailed-balance  and the
beyond-detailed-balance formulations from the observational point
of view. Finally,  section \ref{conclusions} is devoted to the
summary of the obtained results.

\section{Ho\v{r}ava-Lifshitz cosmology}
\label{model}

Let us present the  scenario where the cosmological evolution is
governed by Ho\v{r}ava-Lifshitz gravity
\cite{Calcagni:2009ar,Kiritsis:2009sh}. The dynamical variables
are the lapse and shift functions, $N$ and $N_i$ respectively, and
the spatial metric $g_{ij}$ (roman letters indicate spatial
indices). In terms of these fields the full metric is written as:
\begin{eqnarray}
ds^2 = - N^2 dt^2 + g_{ij} (dx^i + N^i dt ) ( dx^j + N^j dt ) ,
\end{eqnarray}
where indices are raised and lowered using $g_{ij}$. The scaling
transformation of the coordinates reads:
\begin{eqnarray}
 t \rightarrow l^3 t~~~{\rm and}\ \ x^i \rightarrow l x^i~.
\end{eqnarray}

\subsection{Detailed Balance}

The gravitational action is decomposed into a kinetic and a
potential part as $S_g = \int dt d^3x \sqrt{g} N ({\cal L}_K+{\cal
L}_V)$. The assumption of detailed balance \cite{hor3}
  reduces the possible terms in the Lagrangian, and it allows
for a quantum inheritance principle \cite{hor2}, since the
$(D+1)$-dimensional theory acquires the renormalization properties
of the $D$-dimensional one. Under the detailed balance condition
 the full action of Ho\v{r}ava-Lifshitz gravity is given by
\begin{eqnarray}
 S_g &=&  \int dt d^3x \sqrt{g} N \left\{
\frac{2}{\kappa^2}
(K_{ij}K^{ij} - \lambda K^2) \ \ \ \ \ \ \ \ \ \ \ \ \ \ \ \ \  \right. \nonumber \\
&+&\left.\frac{\kappa^2}{2 w^4} C_{ij}C^{ij}
 -\frac{\kappa^2 \mu}{2 w^2}
\frac{\epsilon^{ijk}}{\sqrt{g}} R_{il} \nabla_j R^l_k +
\frac{\kappa^2 \mu^2}{8} R_{ij} R^{ij}
     \right. \nonumber \\
&+&\left.    \frac{\kappa^2 \mu^2}{8(1 - 3 \lambda)} \left[
\frac{1 - 4 \lambda}{4} R^2 + \Lambda  R - 3 \Lambda ^2 \right]
\right\}, \label{acct}
\end{eqnarray}
where
\begin{eqnarray}
K_{ij} = \frac{1}{2N} \left( {\dot{g_{ij}}} - \nabla_i N_j -
\nabla_j N_i \right)
\end{eqnarray}
is the extrinsic curvature and
\begin{eqnarray} C^{ij} \, = \, \frac{\epsilon^{ijk}}{\sqrt{g}} \nabla_k
\bigl( R^j_i - \frac{1}{4} R \delta^j_i \bigr)
\end{eqnarray}
the Cotton tensor, and the covariant derivatives are defined with
respect to the spatial metric $g_{ij}$. $\epsilon^{ijk}$ is the
totally antisymmetric unit tensor, $\lambda$ is a dimensionless
constant and the variables $\kappa$, $w$ and $\mu$ are constants
with mass dimensions $-1$, $0$ and $1$, respectively. Finally, we
mention that in action (\ref{acct}) we have performed the usual
analytic continuation of the parameters $\mu$ and $w$ of the
original version of Ho\v{r}ava-Lifshitz gravity, since such a
procedure is required in order to obtain a realistic cosmology
\cite{Lu:2009em,Minamitsuji:2009ii,Wang:2009rw,Park:2009zra}
(although it could fatally affect the gravitational theory
itself). Therefore, in the present work $\Lambda $ is a positive
constant, which as usual is related to the cosmological constant
in the IR limit.

In order to add the matter component (including both dark and
baryonic matter)  in the theory one can follow two equivalent
approaches. The first is to introduce a scalar field
\cite{Calcagni:2009ar,Kiritsis:2009sh} and thus attribute to dark
matter a dynamical behavior, with its energy density $\rho_m$ and
pressure $p_m$ defined through the field kinetic and potential
energy. Although such an approach is theoretically robust, it is
not suitable from the phenomenological point of view since it
requires special matter-potentials in order to acquire an almost
constant matter equation-of-state parameter ($w_m=p_m/\rho_m$) as
it is suggested by observations. In the second approach one adds a
cosmological stress-energy tensor to the gravitational field
equations, by demanding to recover the usual general relativity
formulation in the low-energy limit
\cite{Sotiriou:2009bx,Carloni:2009jc}. Thus, this matter-tensor is
a hydrodynamical approximation with $\rho_m$ and $p_m$ (or
$\rho_m$ and $w_m$) as parameters. Similarly, one can additionally
include the standard-model-radiation component (corresponding to
photons and neutrinos), with the additional parameters $\rho_r$
and $p_r$ (or $\rho_r$ and $w_r$). Such an approach, although not
fundamental, is better for a phenomenological analysis, such the
one performed in this work.

Now, in order to focus on cosmological frameworks, we impose the
so called projectability condition \cite{Charmousis:2009tc} and
use an FRW metric,
\begin{eqnarray}
N=1~,~~g_{ij}=a^2(t)\gamma_{ij}~,~~N^i=0~,
\end{eqnarray}
with
\begin{eqnarray}
\gamma_{ij}dx^idx^j=\frac{dr^2}{1- K r^2}+r^2d\Omega_2^2~,
\end{eqnarray}
where $ K<,=,> 0$ corresponding  to open, flat, and closed
universe respectively (we have adopted the convention of taking
the scale factor $a(t)$ to be dimensionless and the curvature
constant $ K$ to have mass dimension 2). By varying $N$ and
$g_{ij}$, we obtain the equations of motion:
\begin{eqnarray}\label{Fr1fluid}
H^2 &=&
\frac{\kappa^2}{6(3\la-1)}\Big(\rho_m+\rho_r\Big)+\nonumber\\
&+&\frac{\kappa^2}{6(3\la-1)}\left[ \frac{3\kappa^2\mu^2
K^2}{8(3\lambda-1)a^4} +\frac{3\kappa^2\mu^2\Lambda
^2}{8(3\lambda-1)}
 \right]-\nonumber\\
 &-&\frac{\kappa^4\mu^2\Lambda  K}{8(3\lambda-1)^2a^2} \ ,
\end{eqnarray}
\begin{eqnarray}\label{Fr2fluid}
\dot{H}+\frac{3}{2}H^2 &=&
-\frac{\kappa^2}{4(3\la-1)}\Big(w_m\rho_m+w_r\rho_r\Big)-\nonumber\\
&-&\frac{\kappa^2}{4(3\la-1)}\left[\frac{\kappa^2\mu^2
K^2}{8(3\lambda-1)a^4} -\frac{3\kappa^2\mu^2\Lambda
^2}{8(3\lambda-1)}
 \right]-\nonumber\\
 &-&\frac{\kappa^4\mu^2\Lambda  K}{16(3\lambda-1)^2a^2}\ ,
\end{eqnarray}
where we have defined the Hubble parameter as $H\equiv\frac{\dot
a}{a}$. As usual, $\rho_m$ (dark plus baryonic matter) follows the
standard evolution equation
\begin{eqnarray}\label{rhodotfluid}
&&\dot{\rho}_m+3H(\rho_m+p_m)=0,
\end{eqnarray}
while $\rho_r$ (standard-model radiation) follows
\begin{eqnarray}\label{rhodotfluidrad}
&&\dot{\rho}_r+3H(\rho_r+p_r)=0.
\end{eqnarray}

Finally, concerning the dark-energy sector we can define
\begin{equation}\label{rhoDE}
\rho_{DE}\equiv \frac{3\kappa^2\mu^2 K^2}{8(3\lambda-1)a^4}
+\frac{3\kappa^2\mu^2\Lambda ^2}{8(3\lambda-1)}
\end{equation}
\begin{equation}
\label{pDE} p_{DE}\equiv \frac{\kappa^2\mu^2
K^2}{8(3\lambda-1)a^4} -\frac{3\kappa^2\mu^2\Lambda
^2}{8(3\lambda-1)}.
\end{equation}
The term proportional to $a^{-4}$ is the usual ``dark radiation
term'', present in Ho\v{r}ava-Lifshitz cosmology
\cite{Calcagni:2009ar,Kiritsis:2009sh}, while the constant term is
just the explicit cosmological constant. Therefore, in expressions
(\ref{rhoDE}),(\ref{pDE}) we have defined the energy density and
pressure for the effective dark energy, which incorporates the
aforementioned contributions. Finally, note that using
(\ref{rhoDE}),(\ref{pDE}) it is straightforward to see that these
 dark energy quantities satisfy the
standard evolution equation:
\begin{eqnarray}
\label{DEevol} &&\dot{\rho}_{DE}+3H(\rho_{DE}+p_{DE})=0.
\end{eqnarray}

Using the above definitions, we can re-write the Friedmann
equations (\ref{Fr1fluid}),(\ref{Fr2fluid}) in the standard form:
\begin{equation}
\label{Fr1b} H^2 =
\frac{\kappa^2}{6(3\la-1)}\Big[\rho_m+\rho_r+\rho_{DE}\Big]-
\frac{\kappa^4\mu^2\Lambda  K}{8(3\lambda-1)^2a^2}
\end{equation}
\begin{equation}
\label{Fr2b} \dot{H}+\frac{3}{2}H^2 =
-\frac{\kappa^2}{4(3\la-1)}\Big[p_m+p_r+p_{DE}
 \Big]-\frac{\kappa^4\mu^2\Lambda  K}{16(3\lambda-1)^2a^2}.
\end{equation}
Therefore, if we require these expressions to coincide with the
standard Friedmann equations, in units where $c=1$  we set
\cite{Calcagni:2009ar,Kiritsis:2009sh}:
\begin{eqnarray}
G=\frac{\kappa^2}{16\pi(3\lambda-1)}\nonumber\\
\mu^2\Lambda=\frac{1}{32\pi^2G^2}, \label{simpleconstants0}
\end{eqnarray}
with $G$ the usual Newton's constant. Note that the running of the
light speed with $\lambda$, is not a problem, since in this work
we will remain in the phenomenologically relevant case of
$\lambda=1$.

\subsection{Beyond Detailed Balance}

The aforementioned formulation of Ho\v{r}ava-Lifshitz cosmology
has been performed under the imposition of the detailed-balance
condition. However, in the literature there is a discussion
whether this condition leads to reliable results or if it is able
to reveal the full information of Ho\v{r}ava-Lifshitz
 gravity \cite{Calcagni:2009ar,Kiritsis:2009sh}. Therefore, one
 needs to investigate also the Friedmann equations in the case
 where detailed balance is relaxed. In such a case one can in
 general write
 \cite{Charmousis:2009tc,Sotiriou:2009bx,Bogdanos:2009uj,Carloni:2009jc,Leon:2009rc}:
\begin{eqnarray}\label{Fr1c}
H^2 &=&
\frac{2\sigma_0}{(3\la-1)}\Big(\rho_m+\rho_r\Big)+\nonumber\\
&+&\frac{2}{(3\la-1)}\left[ \frac{\sigma_1}{6}+\frac{\sigma_3
K^2}{6a^4} +\frac{\sigma_4 K}{6a^6}
 \right]+\nonumber\\&+&\frac{\sigma_2}{3(3\la-1)}\frac{ K}{a^2}
\end{eqnarray}
\begin{eqnarray}\label{Fr2c}
\dot{H}+\frac{3}{2}H^2 &=&
-\frac{3\sigma_0}{(3\la-1)}\Big(w_m\rho_m+w_r\rho_r\Big)-\nonumber\\
&-&\frac{3}{(3\la-1)}\left[ -\frac{\sigma_1}{6}+\frac{\sigma_3
K^2}{18a^4} +\frac{\sigma_4 K}{6a^6}
 \right]+\nonumber\\&+&
 \frac{\sigma_2}{6(3\la-1)}\frac{ K}{a^2},
\end{eqnarray}
where $\sigma_0\equiv \kappa^2/12$, and the constants $\sigma_i$
are arbitrary (with $\sigma_2$ being negative). Note that one
could absorb the factor of $6$ in redefined parameters, but we
prefer to keep it in order to coincide with the notation of
\cite{Sotiriou:2009bx,Carloni:2009jc}. As we observe, the effect
of the detailed-balance relaxation is the decoupling of the
coefficients, together with the appearance of a term proportional
to $a^{-6}$. In this case the corresponding quantities for dark
energy are generalized to
\begin{eqnarray}\label{rhoDEext}
&&\rho_{DE}|_{_\text{non-db}}\equiv
\frac{\sigma_1}{6}+\frac{\sigma_3 K^2}{6a^4} +\frac{\sigma_4
K}{6a^6}
\\
&&\label{pDEext} p_{DE}|_{_\text{non-db}}\equiv
-\frac{\sigma_1}{6}+\frac{\sigma_3 K^2}{18a^4} +\frac{\sigma_4
K}{6a^6}.
\end{eqnarray}
Therefore, it is easy to see that
\begin{eqnarray}\label{rhodotfluidnd}
\dot{\rho}_{DE}|_{_\text{non-db}}+3H(\rho_{DE}|_{_\text{non-db}}+p_{DE}|_{_\text{non-db}})=0.
\end{eqnarray}
 Finally, if we force (\ref{Fr1c}),(\ref{Fr2c}) to coincide with
 the standard Friedmann equations, we result to:
\begin{eqnarray}
&&G=\frac{6\sigma_0}{8\pi(3\lambda-1)}\nonumber\\
&&\sigma_2=-3(3\lambda-1). \label{simpleconstants0nd}
\end{eqnarray}

\section{Observational constraints}
\label{Observational constraints}

Having presented the cosmological equations of a universe governed
by Ho\v{r}ava-Lifshitz gravity, both with and without the
detailed-balance condition, we now proceed to study  the
observational constraints on the model parameters. This is
performed in the following two subsections, for the detailed and
non-detailed balance scenarios separately. We mention that since
the   cosmological observations lie deep inside the IR, in the
following we set the running parameter $\lambda$ to $1$.

\subsection{Constraints on Detailed-Balance scenario}

We work in the usual units suitable for observational comparisons,
namely setting  $8\pi G=1$ (we have already set $c=1$ in order to
obtain (\ref{simpleconstants0})). This allows us to reduce the
parameter space, since in this case (\ref{simpleconstants0}) lead
to:
\begin{eqnarray}
\kappa^2=4\nonumber
\\
 \mu^2\Lambda=2. \label{simpleconstants}
\end{eqnarray}
Inserting these relations into Friedmann equation (\ref{Fr1fluid})
we obtain
\begin{equation}
\label{FriedmanDB}
 H^2=\frac13\Big(\rho_m+\rho_r\Big)+\frac13\left(\frac{3 K^2}{2\Lambda
a^4}+\frac{3\Lambda}{2}\right)-\frac{ K}{a^2}.
 \end{equation}
In terms of the usual density parameters
($\Omega_m\equiv\rho_m/(3H^2)$, $\Omega_ K\equiv - K/(H^2a^2)$,
$\Omega_r\equiv\rho_r/(3H^2)$) this expression becomes:
\begin{equation}
 1-\Omega_m-\Omega_r-\Omega_{ K}=\frac{1}{
H^2}\left(\frac{ K^2}{2\Lambda a^4}+\frac{\Lambda}{2}\right).
 \end{equation}
Finally, applying this relation at present time and setting the
current scale factor $a_0=1$ we get:
\begin{equation}
\label{Fr0detb}
 1-\Omega_{m0}-\Omega_{r0}-\Omega_{ K0}=\frac{1}{
H_0^2}\left(\frac{ K^2}{2\Lambda }+\frac{\Lambda}{2}\right),
 \end{equation}
where a $0$-subscript denotes the present value of the
corresponding quantity.

As was mentioned above, we have used the analytic continuation, as
a result of which $\Lambda$ is positive. Thus, relation
(\ref{Fr0detb}) can in principle be satisfied by a suitable choice
of $\Lambda$. However, note that without the analytic continuation
(and therefore with a negative $\Lambda$)   relation
(\ref{Fr0detb}) could never be satisfied (as expected, since in
this case the theory would not have the $\lambda=1$ IR,
general-relativity limit), and this offers another indication,
from the phenomenological point of view, for the necessity of the
analytic continuation in the detailed-balance version of
Ho\v{r}ava-Lifshitz cosmology.

In order to proceed to the elaboration of observational data,  we
consider as usual the matter (dark plus baryonic) component to be
dust, that is $w_m\approx0$, and similarly for the standard-model
radiation we consider $w_r=1/3$, where both assumptions are valid
in the epochs in which observations focus. Therefore, the
corresponding evolution equations
(\ref{rhodotfluid}),(\ref{rhodotfluidrad}) give
$\rho_m=\rho_{m0}/a^3$ and $\rho_r=\rho_{r0}/a^4$ respectively.
Finally, instead of the scale factor it proves convenient to use
the redshift $z$ as the independent variable, which is given by
$1+z\equiv a_0/a=1/a$. Inserting these into Friedmann equation
(\ref{FriedmanDB}) we obtain
\begin{eqnarray}
H^2&=&H_{0}^2\Big\{\Omega_{m0}(1+z)^3+\Omega_{r0}(1+z)^4+\Omega_{ K0}(1+z)^2+\nonumber\\
&\ &+\Big[\omega+\frac{\Omega_{ K 0}^2}{4\omega}(1+z)^4\Big]
\Big\},
 \label{Frdbfinal}
\end{eqnarray}
  where we have also introduced  the
dimensionless parameter $\omega\equiv\Lambda/(2 H_0^2)$. Thus, the
constraint (\ref{Fr0detb}) can be rewritten as:
  \be
\label{cond1}
\Omega_{m0}+\Omega_{r0}+\Omega_{K0}+\omega+\frac{\Omega_{K0}^2}{4\omega}=1.
 \ee

As we have already mentioned above, the term $\Omega_{ K
0}^2/(4\omega)$ is the coefficient of the dark radiation term,
which is a characteristic feature of the Ho\v{r}ava-Lifshitz
gravitational background. Since this dark radiation component has
been present also during the time of nucleosynthesis, it is
subject to bounds from Big Bang Nucleosynthesis (BBN). As
discussed in more details in the Appendix,  if the upper limit on
the total amount of dark radiation allowed during BBN is expressed
through the parameter $\Delta N_\nu$ of the effective neutrino
species \cite{BBNrefs,BBNrefs1,BBNrefs2,Malaney:1993ah}, then we
obtain the following constraint :
  \be
\label{cond2}
 \frac{\Omega_{ K 0}^2}{4\omega}=0.135\dn \Omega_{r0}.
  \ee
Finally, we mention that as usual, the density parameter for
standard model radiation (photons and three species of neutrinos)
$\Omega_{r0}$ is entirely determined by $\Omega_{m0}$, $H_0$ and
the measured value of the CMB temperature \cite{Komatsu:2008hk}.

In most studies of dark energy models it is customary to ignore
curvature (e.g.\cite{DaveCaldwellSteinhardt, LiddleScherrer,
Dutta,Dutta1,Dutta2,Dutta3,Dutta4,Dutta5}), especially concerning
observational constraints. This practice is well motivated for at
least two reasons. Firstly, most inflationary scenarios predict a
high degree of spatial flatness. Secondly, the CMB data impose
stringent constraints on spatial flatness in the context of
constant-$w$ models (for example a combination of WMAP+BAO+SNIa
data \cite{Komatsu:2008hk} provides the tight simultaneous
constraints $-0.0179\leq\Omega_{K0}\leq0.0081$ and
$-0.12\leq1+w\leq0.14$, both at 95\% confidence).

However, it is important to keep in mind that owing to
degeneracies in the CMB power spectrum (see \cite{crooks} and
references therein), the limits on curvature depend on assumptions
regarding the underlying dark energy scenario. For example, if
instead of a constant $w$  one works with a linearly varying $w$,
parameterized as $w\(a\)=w_0+\(1-a\)w_a$, the error on
$\Omega_{K0}$ is much larger, on the order of a few percent
\cite{Wang2,Verde,Ichi1}. The constraints on curvature for
different parameterizations was studied in
\cite{Wright,Ichi2,Ichi3}. The authors of \cite{Ichi3} showed that
for some models of dark energy the constraint on the curvature is
at the level of $5\%$ around a flat universe, whereas for others
the data are consistent with an open universe with
$\Omega_{K0}\sim0.2$. According to \cite{Verde}, geometrical tests
such as the combination of the Hubble parameter $H(z)$ and the
angular diameter distance $D_A(z)$, using (future) data up to
sufficiently high redshifts $z\sim 2$, might be able to
disentangle curvature from dark energy evolution, though not in a
model-independent way. \cite{Cortes,Virey} highlight the pitfalls
arising from ignoring curvature in studies of dynamical dark
energy, and recommend treating $\Omega_{K0}$ as a free parameter
to be fitted along with the other model parameters.

In the present work,  the spatial curvature  plays a very crucial
role, since, as it has been extensively stated in the literature
\cite{Calcagni:2009ar,Kiritsis:2009sh},  Ho\v{r}ava-Lifshitz
cosmology coincides completely with $\Lambda$CDM if one ignores
curvature. Therefore, and following the discussion above, we
choose to treat $\Omega_{K0}$ as a free parameter.

In summary, the scenario at hand involves four parameters (we fix
$H_0$ by its 5-year WMAP best-fit values, given in Table 1 of
\cite{Komatsu:2008hk}), namely $\Omega_{m0}$, $\Omega_{K0}$,
$\omega$ and $\dn$, subject to constraint  equations (\ref{cond1})
and (\ref{cond2}). Therefore, only two of these parameters are
independent. Although  one would usually expect to be able to
choose two of them at will, the non-linear nature of the
constraint equations does not facilitate this, and one has no
choice but to use
 $\Omega_{m0}$ and $\dn$  as free parameters.
Inverting (\ref{cond1}) and (\ref{cond2}) to express $\omega$ and
$\Omega_{K0}$ in terms of these independent parameters
  for a given curvature, we obtain:
    \ba
     \label{invert1}
\omega\(K;\Omega_{m0},\dn\)=1-\Omega_{m0}-(1-\dn)\Omega_{r0}\nonumber&&\\
-0.73\text{
sgn}\(K\)\sqrt{\dn}\sqrt{\Omega_{r0}-\Omega_{m0}\Omega_{r0}-\Omega_{r0}^2}&&
\ea
  and
      \be \label{invert2}
\vert\Omega_{K0}\(\Omega_{m0},\dn\)\vert=\sqrt{0.54\,\dn\,\Omega_{r0}\,\omega\(\Omega_{m0},\dn\)}.
\ee

The BBN upper limit on $\dn$ is $-1.7\leq\dn\leq2.0$
\cite{BBNrefs,BBNrefs1,BBNrefs2,Malaney:1993ah}. A negative value
of $\dn$ (which is usually associated with models involving decay
of a massive particle) is  not possible in the present  model,
since $\omega$ (i.e. $\Lambda$) is always positive. Additionally,
$\dn=0$ corresponds to the zero curvature scenario (a
non-interesting case since
 Ho\v{r}ava-Lifshitz cosmology with zero curvature becomes
indistinguishable from $\Lambda$CDM).

\begin{figure}[ht]
\begin{center}
\includegraphics[width=8cm]{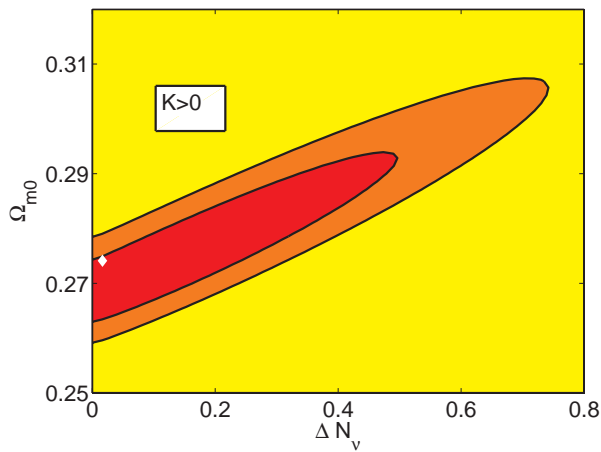}
\caption{(Color Online) {\it{ Contour plots of $\Omega_{m0}$ vs
        $\dn$ for   positive curvature ($K>0$), under SNIa, BAO and CMB observational data.
        The yellow (light) region
is excluded at the 2$\sigma$ level, and the orange (darker) region
is excluded at the 1$\sigma$ level.  The red (darkest) region is
not excluded at either confidence level. The white diamond marks
the best-fit point. The model parameters $\omega\equiv\Lambda/(2
H_0^2)$ and $\Omega_{K0}\equiv - K/(H^2_0)$ are related to
$\Omega_{m0}$ and $\dn$ through  (\ref{invert1}) and
(\ref{invert2}). }}} \label{posdb}
\end{center}
\end{figure}

\begin{figure}[ht]
\begin{center}
\includegraphics[width=8cm]{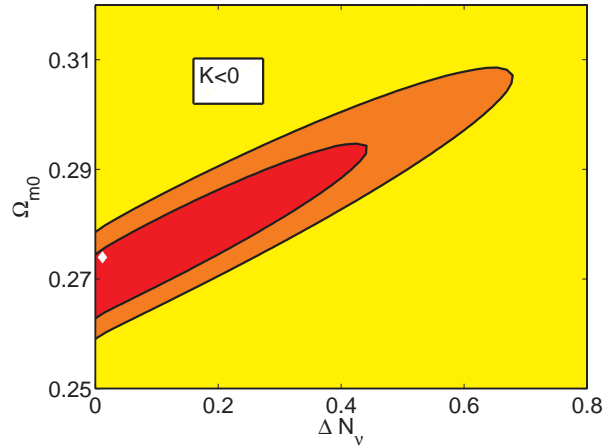}
\caption{ (Color Online) {\it{Contour plots of $\Omega_{m0}$ vs
        $\dn$ for negative curvature ($K<0$), under SNIa, BAO and CMB observational data.
           The yellow (light) region
is excluded at the 2$\sigma$ level, and the orange (darker) region
is excluded at the 1$\sigma$ level.  The red  region is not
excluded at either confidence level. The white diamond marks the
best-fit point. The model parameters  $\omega\equiv\Lambda/(2
H_0^2)$ and $\Omega_{K0}\equiv - K/(H^2_0)$ are related to
$\Omega_{m0}$ and $\dn$ through  (\ref{invert1}) and
(\ref{invert2}). }}} \label{negdb}
\end{center}
\end{figure}

In Fig. \ref{posdb} and Fig \ref{negdb} we use a combination of
observational data from SNIa, BAO and CMB to construct  likelihood
contours for the parameters $\Omega_{m0}$ and $\dn$ for positive
and negative curvatures respectively.
These figures show that the Ho\v{r}ava-Lifshitz cosmological
scenario under the detailed balance condition is not ruled out by
observations. However, they lead to tight constraints on the
amount of dark radiation allowed at the time of nucleosysnthesis
(tighter than the corresponding limits from BBN), and thus to the
parameter $\Lambda$.  For example, the $1\sigma$ limits on
 $\Omega_{K0}$, $\Lambda$ and $\mu$ (which is actually connected
 to $\Lambda$ through (\ref{simpleconstants0}))  are presented in Table \ref{dblimits}.
For convenience we have kept the factors of $H_0$ and $8\pi G$.
Thus, one can either use the usual ansatz $8\pi G=H_0=1$, or
insert physical units using $H_0=1.503 \times 10^{-42}\,$GeV and
$8\pi G=1.681\times10^{-37}\,$ GeV$^{-2}$. In the later case, one
obtains $0<\Lambda\lesssim 1.86 \times10^{-9}\,$ eV$^4$ and
$3.30\times 10^{60}\lesssim\mu<\infty$ for the positive curvature
case, and similarly for the negative curvature one.
\begin{table*}
    \centering
        \begin{tabular}{|c|c|c|c|}
        \hline $\kappa^2/(8\pi G)$ &
        \textbf{$\Omega_{K0}$} & \textbf{$\left(8\pi G/H_0^2\right)\Lambda $}& $ \left(H_0\sqrt{8\pi G}\right)\mu$\\\hline
        4&     $(0,\, 0.0038)$       & $(0,\,1.4189 )$ & $(1.1872,\infty)$ \\\hline
           4&   $(-0.0039,\,0)$       & $(0,\,1.4063) $ & $(1.1925,\infty)$ \\\hline
        \end{tabular}
    \caption{1$\sigma$ limits on the parameter values for the detailed-balance scenario, for positive and negative curvature.}
    \label{dblimits}
\end{table*}

In conclusion, we have shown that  Ho\v{r}ava-Lifshitz cosmology
under the assumption of detailed-balance condition cannot fulfill
observational requirements without the analytic continuation
transformation. Under analytic continuation the observational
constraints on the parameters are quite tight. This feature was
already mentioned in \cite{Bogdanos:2009uj}, following qualitative
theoretical arguments concerning the effective light speed in
Ho\v{r}ava-Lifshitz framework, where it was stated that a  fine
tuning would be needed as a way out. The analysis of this section
offers  new, phenomenological indications towards the direction of
tight constraints.

\subsection{Constraints on Beyond-Detailed-Balance scenario}

In units where $8\pi G=1$ relations (\ref{simpleconstants0nd})
give
\begin{eqnarray}
&&\sigma_0=1/3\nonumber\\
&&\sigma_2=-6. \label{simpleconstantsnd}
\end{eqnarray}
Using these values and following the procedure of the previous
subsection, the Friedmann equation (\ref{Fr1c}) can be written as
 \ba
   \label{Fr1c_a}
H^2&=&H_{0}^2\Big\{\Omega_{m0}\(1+z\)^3+\Omega_{r0}\(1+z\)^4+\Omega_{ K0}\(1+z\)^2+\nonumber\\
&\ &+\Big[\omega_1+\omega_3 \(1+z\)^4+\omega_4
\(1+z\)^6\Big]\Big\}.
  \ea
  In this expression  we have introduced the dimensionless
 parameters $\w_1$, $\w_3$ and $\w_4$, related to the model parameters $\sigma_1$, $\sigma_3$ and $\sigma_4$
 through:
\ba
\omega_1&=&\frac{\sigma_1}{6H_0^2}\nonumber\\
\omega_3&=&\frac{\sigma_3 H_0^2\Omega_{ K0}^2}{6}\nonumber\\
\omega_4&=&-\frac{\sigma_4\Omega_{ K0}}{6}.
  \ea
   Furthermore, we consider the
combination $\omega_4$ to be positive, in order to ensure that the
Hubble parameter is real for all redshifts. $\omega_4>0$ is
required also for the stability of the gravitational perturbations
of the theory \cite{Sotiriou:2009bx,Bogdanos:2009uj}. For
convenience we moreover assume $\sigma_3\geq0$, that is
$\omega_3\geq0$.

The scenario at hand involves the following parameters:
  $H_0$, $\Omega_{m0}$, $\Omega_{K0}$, $\w_1$, $\w_3$
and $\omega_4$. Similarly to the detailed-balance section these
are subject to two constraints. The first one arises from the
Friedman equation at $z=0$, which leads to
  \be
     \label{ndbcond1}
\Omega_{m0}+\Omega_{r0}+\Omega_{K0}+\w_1+\w_3+\w_4=1.
  \ee
  This constraint eliminates the parameter $w_1$.

 The
second constraint arises from BBN considerations. The term
involving $\w_3$ represents the usual dark-radiation component. In
addition, the $\w_4$-term represents a kination-like component (a
quintessence field dominated by kinetic energy
\cite{kination,kination1}). If $\dn$ represents the BBN upper
limit on the total energy density of the universe beyond standard
model constituents, then as we show in the Appendix we acquire the
following constraint at the time of BBN ($z=z_{\rm BBN}$)
\cite{BBNrefs,BBNrefs1,BBNrefs2,Malaney:1993ah}:
 \be
\label{ndbcond2} \w_3+\w_4\(1+z_{\rm BBN}\)^2=0.135\dn\Omega_{r0}.
\ee
  It is clear that BBN imposes an extremely strong constraint on
$\w_4$, since its largest possible value (corresponding to
$\w_3=0$) is $\sim 10^{-24}$.
\begin{figure}[!]
\begin{center}
\includegraphics[width=8cm]{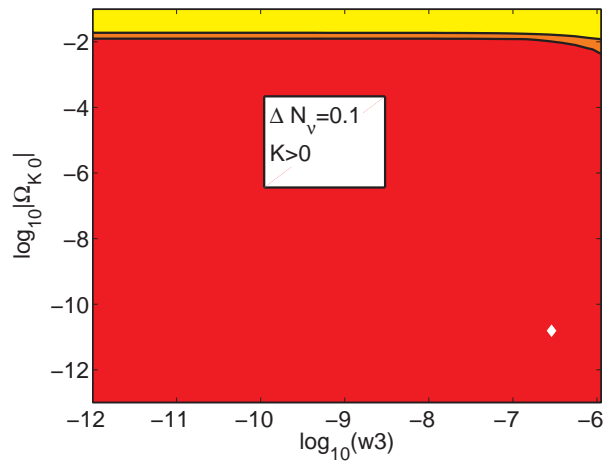}
\caption{ (Color Online) {\it{Contour  plots of $\log_{10}(w_3)$
vs $\log_{10}\vert\Omega_{ K0}\vert$ for $ K>0$ and $\dn=0.1$,
using SNIa, BAO and CMB data.     The yellow (light) region is
excluded at the 2$\sigma$ level, and the orange (darker) region is
excluded at the 1$\sigma$ level.  The red (darkest)  region is not
excluded at either confidence level.  The white diamond marks the
best-fit point. }}} \label{posk0p1}
\end{center}
\end{figure}
\begin{figure}[!]
\begin{center}
\includegraphics[width=8cm]{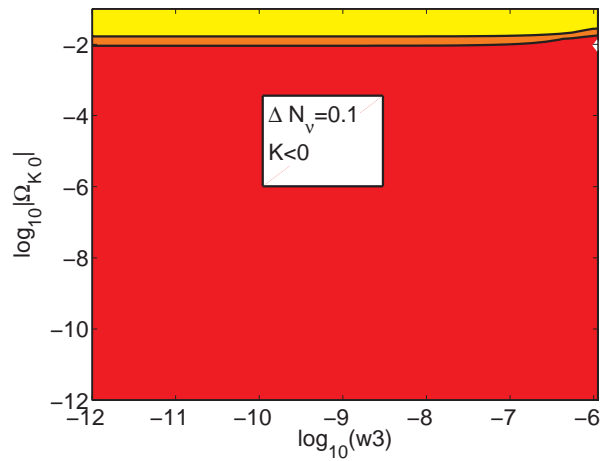}
\caption{ (Color Online) {\it{Contour  plots of $\log_{10}(w_3)$
vs $\log_{10}\vert\Omega_{ K0}\vert$ for $ K<0$ and $\dn=0.1$,
  using SNIa, BAO and CMB data.     The yellow (light) region
is excluded at the 2$\sigma$ level, and the orange (darker) region
is excluded at the 1$\sigma$ level.  The red (darkest) region is
not excluded at either confidence level. The white diamond marks
the best-fit point (near the top right corner in this case).}}}
\label{negk0p1}
\end{center}
\end{figure}

We use relation (\ref{ndbcond2}) to eliminate $\w_4$ in favor of
$\w_3$ and $\dn$, and treat $\w_3$ and $\Omega_{K0}$ as our free
parameters. Since $\omega_4$ is non-negative,  relation
(\ref{ndbcond2}) determines also the upper bound of $\omega_3$.
For the remaining parameters, $\Omega_{m0}$ and $H_0$
  (unless otherwise stated) we assume priors
corresponding to their 5-year WMAP best-fit values (given in Table
1 of \cite{Komatsu:2008hk}).

\begin{figure}[ht]
\begin{center}
\includegraphics[width=8cm]{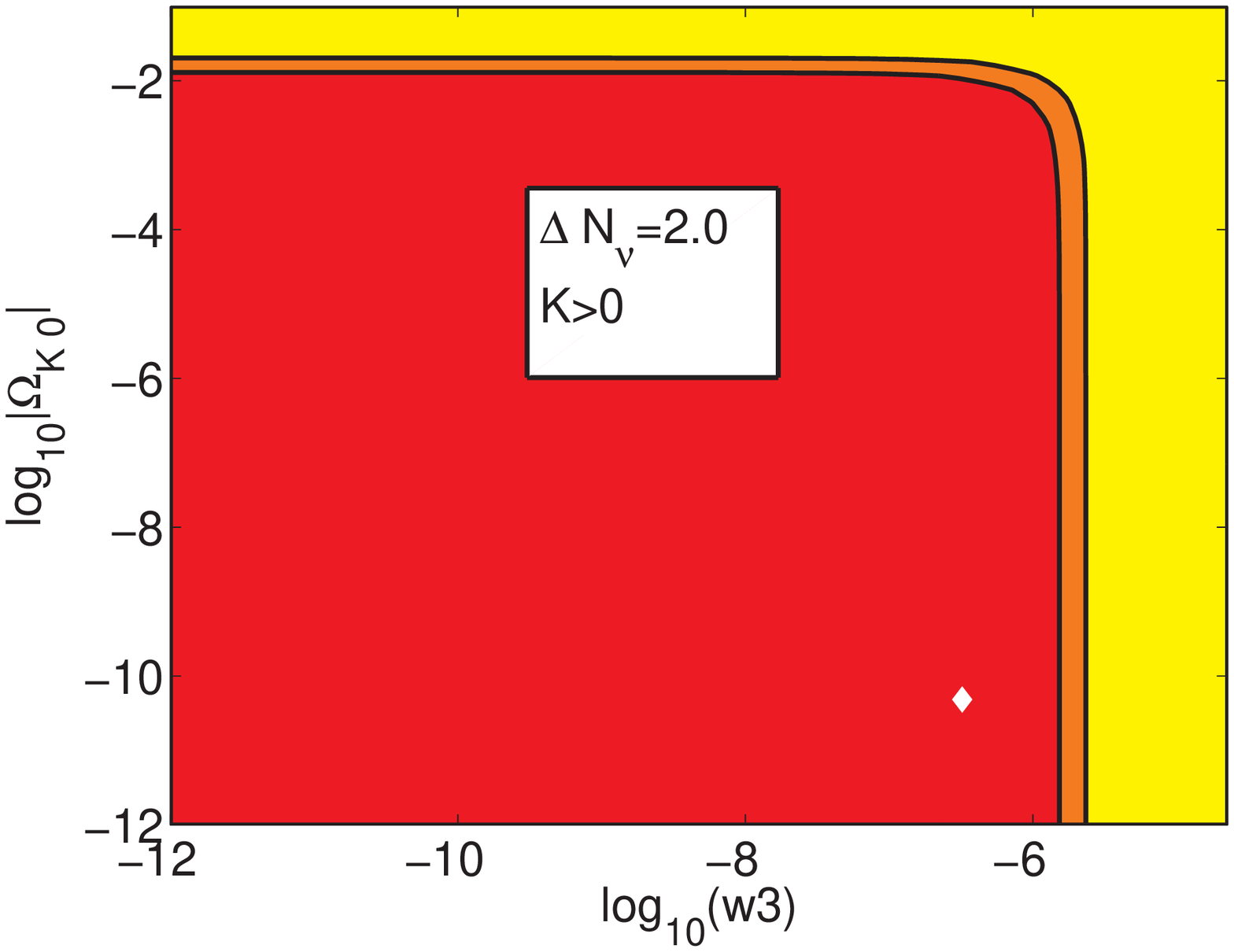}
\caption{ (Color Online) {\it{Contour  plots of $\log_{10}(w_3)$
vs  $\log_{10}\vert\Omega_{ K0}\vert$ for $ K>0$ and $\dn=2.0$,
using SNIa, BAO and CMB data.     The yellow (light) region is
excluded at the 2$\sigma$ level, and the orange (darker) region is
excluded at the 1$\sigma$ level.  The red (darkest) region is not
excluded at either confidence level. The white diamond marks the
best-fit point. }}} \label{posk2}
\end{center}
\end{figure}
\begin{figure}[ht]
\begin{center}
\includegraphics[width=8cm]{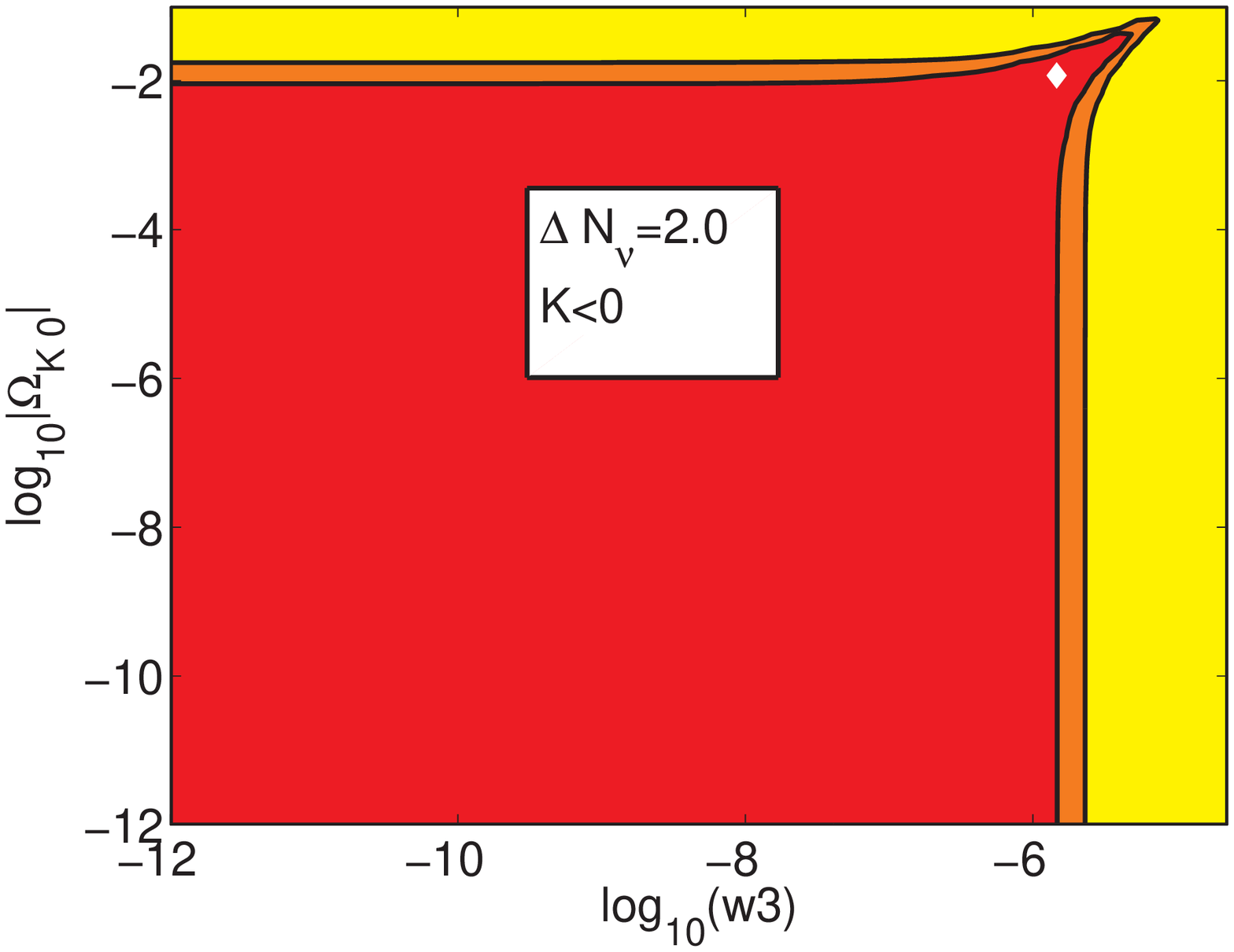}
\caption{ (Color Online) {\it{Contour  plots of $\log_{10}(w_3)$
vs $\log_{10}\vert\Omega_{ K0}\vert$ for $ K<0$ and $\dn=2.0$,
using SNIa, BAO and CMB data.     The yellow (light) region is
excluded at the 2$\sigma$ level, and the orange (darker) region is
excluded at the 1$\sigma$ level.  The red (darkest) region is not
excluded at either confidence level. The white diamond marks the
best-fit point.}}} \label{negk2}
\end{center}
\end{figure}

We now proceed to constrain  the free parameters $\Omega_{ K0}$
and $\omega_3$ through observations.  Using the combined
SNIa+CMB+BAO data, we construct likelihood contours for these two
parameters. The details and the techniques of the construction are
presented in the Appendix. Furthermore, since the BBN limits on
$\dn$ vary depending on assumptions, in addition to our canonical
choice of upper limit $\dn=2.0$, we have also considered the more
restrictive limit of $\dn=0.1$.
\begin{table*}
    \centering
        \begin{tabular}{|c|c|c|c|c|c|c|}
        \hline
        $\sigma_0/(8\pi G)$&    \textbf{$\dn$} &  \textbf{$\Omega_{K0}$}        & \textbf{$\left(8\pi G/H_0^2\right)\sigma_1$} & $\sigma_2$       & \textbf{$\(8\pi G H_0^2\)\sigma_3$}          &  \textbf{$\sigma_4/(8\pi G)$}              \\\hline
       1/3& $0.1$          &     $(0,\, 0.01)   $          &   $(4.29,\,4.33)$ & $-6$     &         $(0,\, 0.03)$   &          $(-9.08\times 10^{-22},\,0)$                \\\hline
          1/3& $0.1$          &     $(-0.01,\, 0)   $          &   $(4.40,\,4.45)$ & $-6$      &         $(0,\, 0.81)$   &          $(0,\,5.66\times 10^{-22})$                \\\hline
         1/3& $2.0$          &     $(0,\, 0.04)   $          &   $(4.13,\,4.45)$ & $-6$      &         $(0,\, 0.01)$   &          $(-1.77\times 10^{-20},\,-2.62\times 10^{-21})$    \\\hline
          1/3&  $2.0$          &     $(-0.01,\, 0)   $          &   $(4.40,\,4.45)$  & $-6$    &         $(0,\, 0.23)$   &          $(-2.61\times 10^{-20},\,-1.16\times 10^{-20})$                \\\hline
        \end{tabular}
    \caption {1$\sigma$ limits on the parameter
     values for the beyond-detailed-balance scenario, for positive
     and negative curvature, and for two values of the effective neutrino
species parameter $\Delta N_\nu$ (see text).}
    \label{ndblimits}
\end{table*}

Figures \ref{posk0p1} and \ref{negk0p1} depict the $1\sigma$ and
$2\sigma$ $\w_3-\vert\Omega_{ K 0}\vert$ contours, for $\dn=0.1$,
for positive and negative curvature respectively. Figures
\ref{posk2} and \ref{negk2} are the corresponding plots using
$\dn=2$.
 In each case, $\w_3$
extends over its entire allowed range, namely
$0\leq\w_3\leq0.135\dn\Omega_{r0}$.

As we observe, the Ho\v{r}ava-Lifshitz cosmological scenario
beyond the detailed balance condition is not ruled out by
observations. However, they impose strong constraints  on $\w_3$
(for the case of $\dn=2.0$ the constraints on $\w_3$ for both the
positive and negative curvature cases are stronger than the upper
bound from BBN), and extremely tight constraints on $\omega_4$.
Additionally, the constraints on the curvature are of the order of
a percent. Note that the contours expand as $\dn$ is reduced. This
is  expected since the smaller the amount of exotic components
(dark-radiation and kination-like ones), the closer the model is
to $\Lambda$CDM.

The approximate $1\sigma$ limits on the model parameters
$\sigma_i$ are presented in Table \ref{ndblimits}. The lower limit
on $\sigma_3$ is zero. From (\ref{ndbcond1}) and (\ref{ndbcond2})
it is clear that $\sigma_1$ and $\sigma_4$ attain their maximum
values when $\sigma_3$ is at its minimum, and vice versa.
Similarly to the previous subsection, one can either use the usual
ansatz $8\pi G=H_0=1$, or insert physical units using $H_0=1.503
\times 10^{-42}\,$GeV and $8\pi G=1.681\times10^{-37}\,$
GeV$^{-2}$. In the later case, one obtains that $\sigma_1$ is
tightly constrained to be at the level of the cosmological
constant $(10^{-12}\,$eV$^4)$, as expected. Additionally, the data
impose extremely stringent constraints on $\sigma_4$, which was
also expected. However, even for such small values, the
phenomenological implications of the kination-like
$\sigma_4$-component are very interesting. As discussed in detail
in \cite{Salati, Pallis}, it could dominate the universe prior to
BBN and it could significantly affect the freeze-out, and hence
the relic abundances  of neutralino  dark matter, by a few orders
of magnitude. For dark matter that decays into leptons (see e.g.
\cite{Ma}) this could be relevant to recent observations of high
energy positrons and electrons by the PAMELA
\cite{pamela1,pamela2} and ATIC \cite{atic} experiments.

\section{Conclusions}
\label{conclusions}

In this work we constrained Ho\v{r}ava-Lifshitz cosmology using
observational data. In particular, we considered scenarios where
the gravitational sector is forced to satisfy the detailed-balance
condition, and also  those where this condition is relaxed.
Additionally, we have included the matter and radiation sectors
following the usual effective fluid approach. These constructions,
which cover the range of Ho\v{r}ava-Lifshitz cosmology, were
confronted with data from BBN, SNIa, CMB and BAO observations.

Our first result is that the detailed-balance formulation of
Ho\v{r}ava-Lifshitz gravity cannot fulfill observational
requirements, without the analytic-continuation transformation.
Under the analytic continuation we found that Ho\v{r}ava-Lifshitz
cosmology can be compatible with observations, and we presented
the corresponding contour-plots on the model parameters. These
likelihood-contours impose tight constraints on the model
parameters, and the corresponding 1$\sigma$-bounds are presented
in Table \ref{dblimits}. However, we mention that although
analytic continuation is necessary for a realistic cosmology, it
can fatally affect the gravitational theory itself, spoiling its
initial stability and well-behaving nature. Therefore, the
detailed-balance version of Ho\v{r}ava-Lifshitz cosmology seems
rather unlikely to be a robust description of nature.

The version of Ho\v{r}ava-Lifshitz cosmology in which the
detailed-balance condition has been abandoned, is also compatible
with observations. We constructed the likelihood-contours for the
two involved free parameters, namely the curvature and the
dark-radiation coefficients. As we showed, observations lead to
strong bounds in these parameters, and the corresponding
1$\sigma$-allowed ranges are presented in Table \ref{ndblimits}.
This  feature was expected, since the data refer to redshifts in
which the novel terms of Ho\v{r}ava-Lifshitz cosmology are
downgraded. However, these terms can have significant cosmological
implications prior to nucleosynthesis, which could be probed by
recent observations of high energy positrons and electrons by the
PAMELA \cite{pamela1,pamela2} and ATIC \cite{atic} experiments.

Although the present analysis indicates that Ho\v{r}ava-Lifshitz
cosmology can be compatible with observations, it does not
enlighten the discussion about possible conceptual problems and
instabilities of Ho\v{r}ava-Lifshitz gravity,  nor it can
interfere with the questions concerning the validity of its
theoretical background, which is the subject of interest of other
studies. In particular, without a solid theoretical basis, it is
not clear whether Ho\v{r}ava-Lifshitz gravity is able to pass the
basic parametrized post newtonian (PPN) tests that any physically
interesting gravitational theory should \cite{PPN,PPN1,PPN2,PPN3}.
The present work just faces the problem from the phenomenological
point of view, and thus its results can been taken into account
only if Ho\v{r}ava-Lifshitz  gravity passes successfully the
aforementioned theoretical tests.

\begin{acknowledgments}
The authors would like to thank Bob Scherrer for useful
discussions.
\end{acknowledgments}

\appendix*

\section{Observational data and constraints}
\label{Observational data and constraints}

In this appendix,  we briefly review the main sources of
observational constraints used in this work, namely, Big Bang
Nucleosynthesis (BBN), Baryon Acoustic Oscillations (BAO) and the
Cosmic Microwave Background (CMB).\\

{\it{a. Big Bang Nucleosynthesis constraints}}\\

Big Bang Nucleosynthesis (BBN)  provides  a highly sensitive tool
for probing physics beyond the standard model (for reviews see
e.g. \cite{BBNrefs,BBNrefs1,BBNrefs2,Malaney:1993ah}). Abundances
of light elements predicted by BBN, particularly the $^4$He one,
are sensitive to the expansion rate of the universe (or
equivalently to its total energy density) at the time of BBN.
Additionally, the abundances depend also on the baryon to photon
ratio, though this ratio can be independently determined from the
CMB by WMAP data \cite{Komatsu:2008hk}. BBN therefore imposes
constraints on the densities of possible extra exotic radiation
constituents (beyond the standard model photons and three flavors
of neutrinos).

The constraints on  the energy density of these exotic
constituents are usually expressed in terms of the effective
neutrino species $\Delta N_{\nu}$. Assuming that neutrinos are
fully decoupled from photons and do not gain energy from $e^{\pm}$
annihilation, they are colder than photons by a factor of
$T_\nu/T_\gamma=\(4/11\)^{1/3}$ (see e.g. \cite{BBNrefs2}). Using
also that the neutrino energy is related to the photon one by a
factor of $7/8$, the total energy density of relativistic species
- photons and $(3+\Delta N_\nu)$ species of neutrinos -  reads
 \be \rho_{T}=\rho_{\gamma}+\(3+\Delta
N_\nu\)\(\frac78\)\(\frac{4}{11}\)^{4/3}\rho_\gamma.
 \ee
  In terms
of the total standard-model relativistic density (photons and
three species of neutrinos) $\rho_{\gamma\nu}$, the above can be
written as
\be
 \rho_T=\(1+0.135\Delta N_{\nu}\)\rho_{\gamma\nu}.
\ee

In the present work  we  use the upper limits on $\Delta N_{\nu}$
provided in \cite{BBNrefs2}:
 $-1.7\leq\Delta N_{\nu}\leq2.0$, although
more restrictive bounds have also been proposed  (see e.g.
\cite{WalkerZentner,BeanHansenMelchiorri}). We mention that the
above limits do not apply to models in which the dark radiation or
other exotic components are injected later than BBN (see
\cite{DuttaHsu} for an example of such a model), however they are
obviously applicable to Ho\v{r}ava-Lifshitz cosmology, in which
dark radiation is always present, arising from the gravitational
theory itself.
\\

{\it{b. Type Ia Supernovae constraints}}\\

In order to incorporate supernova constraints  we use the Union08
compilation of SnIa data \cite{union08}. This is a heterogeneous
data-set, consisting of data from the Supernova Legacy Survey, the
Essence survey, the recently extended data-set of distant
supernovae observed by the Hubble Space Telescope, as well as
older data-sets.

The $\chi^2$ from SNIa is calculated as:
  \be \chi ^2 _{SN} =
\frac{{\sum\limits_{i = 1}^N {\left[ {\mu _{\text{obs} } \left(
{z_i } \right) - \mu _{\rm th} \left( {z_i } \right)} \right]} ^2
}}{{\sigma^{2} _{\mu,i} }},
 \ee
    where $N=307$ is the number of SNIa
data points. $\mu_{\rm obs}$ is the  observed distance modulus,
defined as the difference between the apparent and absolute
magnitude of the supernova. The $\sigma_{\mu,i}$ are the errors in
the observed distance moduli, arising from a variety of sources,
and assumed to be gaussian and uncorrelated. The theoretical
distance modulus $\mu_{\rm th}$  depends on the model parameters
$a_i$ via the dimensionless luminosity distance $D_{L}(z;a_i)$:
\be D_{L}\(z;a_i\)\equiv\left(1+z\right)
\int^{z}_{0}dz'\frac{H_0}{H\left(z';a_i\right)},
 \ee as follows:
\be \mu_{\rm
th}\left(z\right)=42.38-5\log_{10}h+5\log_{10}\[D_{L}\left(z;a_i\right)\].
\ee
 The marginalization over the present value of the Hubble
 parameter is performed
following the techniques described in \cite{perivol1}, and we
construct $\chi^2$ likelihood contours for the various model
parameters.
\\

{\it{c. CMB constraints}}\\

We use the CMB data to impose constraints on the parameter space,
following the recipe described in \cite{Komatsu:2008hk}.
 The ``CMB
shift parameters'' \cite{Wang1,Wang2} are defined as:
 \be
  R\equiv
\sqrt{\Omega_{m0}}H_0 r\(z_*\),\,\quad l_{a}\equiv \pi
r\(z_*\)/r_{s}\(z_*\).
 \ee
  $R$ can be physically interpreted as a
scaled distance to recombination, and $l_{a}$ can be interpreted
as the angular scale of the sound horizon at recombination. $r(z)$
is the comoving distance to redshift $z$ defined as
 \be
r(z)\equiv\int_{0}^{z}\frac{1}{H\(z\)}dz,
 \ee
 while $r_{s}\(z_*\)$ is
the comoving sound horizon at decoupling (redshift $z_*$), given
by
 \be
r_{s}\(z_*\)=\int_{z_*}^{\infty}\frac{1}{H\(z\)\sqrt{3\(1+R_{b}/\(1+z\)
\)}}dz.
 \ee
  The quantity $R_b$ is the ratio of the energy density
of photons to baryons, and its value can be calculated as
$R_b=31500 \Omega_{b0} h^2 \(T_{CMB}/2.7K\)^{-4}$, ($\Omega_{b0}$ being the present day density parameter for baryons) using
$T_{CMB}=2.725$  \cite{Komatsu:2008hk}. The redshift at decoupling
$z_*\(\Omega_{b0},\Omega_{m0},h\)$ can be calculated from the
following fitting formula \cite{husugiyama}:
 \be
z_*=1048\[1+0.00124\(\Omega_{b0}
h^2\)^{-0.738}\]\[1+g_1\(\Omega_{m0} h^2\)^{g_2}\], \ee with $g1$
and $g2$ given by:
\begin{eqnarray*}
g_1&=&\frac{0.0783\(\Omega_{b0} h^2\)^{-0.238}}{1+39.5\(\Omega_{b0} h^2\)^{0.763}}\\
g_2&=&\frac{0.560}{1+21.1\(\Omega_{b0} h^2\)^{1.81}}.
\end{eqnarray*}
Finally, the $\chi^2$ contribution of the CMB reads
 \be
\chi^{2}_{CMB}=\mathbf{V}_{\rm CMB}^{\mathbf{T}}\mathbf{C}_{\rm
inv}\mathbf{V}_{\rm CMB}. \ee
 Here $\mathbf{V}_{\rm
CMB}\equiv\mathbf{P}-\mathbf{P}_{\rm data}$, where $\mathbf{P}$ is
the vector $\(l_{a},R,z_{*}\)$ and the vector $\mathbf{P}_{\rm
data}$ is formed from the WMAP $5$-year maximum likelihood values
of these quantities \cite{Komatsu:2008hk}. The inverse covariance
matrix $\mathbf{C}_{\rm inv}$ is also provided in
\cite{Komatsu:2008hk}.
\\

{\it{d. Baryon Acoustic Oscillation constraints}}\\

In this case the measured quantity  is the ratio
$d_z=r_{s}\(z_{d}\)/D_{V}\(z\)$, where $D_{V}\(z\)$ is the so
called ``volume distance'', defined in terms of the angular
diameter distance $D_{A}\equiv r\(z\) /\(1+z\)$ as
\be
D_{v}\(z\)\equiv\left[\frac{\(1+z\)^2 D_{A}^{2}(z) z
}{H(z)}\right]^{1/3},
 \ee
and $z_d$ is the redshift of the baryon drag epoch, which can be
calculated from the fitting formula \cite{HuEisenstein}: \be
z_d=\frac{1291\(\Omega_{m0} h^2\)^{0.251}}{1+\(\Omega_{M0}
h^2\)^{0.828}}\[1+b_1\(\Omega_{b0} h^2\)^{b_2}\],
 \ee
where $b_1$ and $b_2$ are given by
\begin{eqnarray*}
b_1&=&0.313\(\Omega_{m0} h^2\)^{-0.419}\[1+0.607\(\Omega_{m0} h^2\)^{0.674}\]\\
b2&=&0.238\(\Omega_{m0} h^2\)^{0.223}.
\end{eqnarray*}

We use the two measurements of $d_z$ at redshifts $z=0.2$ and
$z=0.35$ \cite{Percival:2009xn}. We calculate the $\chi^2$
contribution of the BAO measurements as: \be
\chi^{2}_{BAO}=\mathbf{V}_{\rm BAO}^{\mathbf{T}}\mathbf{C}_{\rm
inv}\mathbf{V}_{\rm BAO}.
 \ee
    Here the vector $\mathbf{V}_{\rm
BAO}\equiv\mathbf{P}-\mathbf{P}_{\rm data}$, with
$\mathbf{P}\equiv \( d_{0.2},d_{0.35} \) $, and $\mathbf{P}_{\rm
data}\equiv\(0.1905, 0.1097\)$, the two measured BAO data points
\cite{Percival:2009xn}. The inverse covariance matrix is provided
in \cite{Percival:2009xn}.


\end{document}